\documentstyle[12pt]{article}
\begin{document}
\title{Dynamical Breakdown of Chirality and Parity\\
       in (2+1)-dimensional QED
}

\vspace{2.0em}

\author{%
  {\sc Kei-Ichi Kondo}%
     \thanks{e-mail: kondo@tansei.cc.u-tokyo.ac.jp;
                     kondo@cuphd.nd.chiba-u.ac.jp},
  {\sc Toru Ebihara}%
     \thanks{e-mail: ebihara@cuphd.nd.chiba-u.ac.jp},
  {\sc Takuya Iizuka}%
     \thanks{e-mail: tiizuka@cuphd.nd.chiba-u.ac.jp},
  \\
  and
  {\sc Eiji Tanaka}%
     \thanks{e-mail: etanaka@cuphd.nd.chiba-u.ac.jp}
  \vspace{1.2em}
  \\
  Department of Physics, Faculty of Science \\
  $\&$ Graduate School of Science and Technology,\\
         Chiba University, Chiba 263, Japan
}

\date{
  {\bf\tt CHIBA-EP-77-REV\\ July  1994\\ hep-ph/9404361}
}

\maketitle

\abstract{%
In the (2+1)-dimensional QED  with and without the
Chern-Simons term, we find the non-local gauge in which
there is no wavefunction renormalization for the fermion
in the framework of the Schwinger-Dyson equation.
By solving the Schwinger-Dyson equation in the non-local
gauge, we find a finite critical value $N_f^c$ for the
number of flavors $N_f$ of four-component Dirac fermions,
above which the chiral symmetry restores irrespective of
presence or absence of the Chern-Simons term.
In the same framework, we study the possibility of
dynamical breakdown of the parity.  It is shown that the
parity is not dynamically broken.  We discuss this reason
from the viewpoint of the Schwinger-Dyson equation.
 }

\newpage

\section{Introduction}
The study of the dynamical symmetry breaking in the
(2+1)-dimensional QED (QED3) has attracted a great deal of attention recently.
In the framework of the Schwinger-Dyson (SD) equation in the
relativistic quantum field theory, Appelquist et al.
\cite{ABKW86a,ANW88} investigated the dynamical symmetry breaking
of $U(2N_f)$ to $U(N_f) \times U(N_f)$ in
QED3 with $N_f$ flavors of four-component Dirac fermions.
The study has given rise to much controversy as follows.
 In the bare vertex
approximation: $\Gamma_\mu(p,q)=\gamma_\mu$, Appelquist, Nash and
Wijewardhana (ANW) \cite{ANW88} have shown that there
exists the finite critical number of flavors
\begin{equation}
 N_f^c = {32 \over \pi^2} \sim 3.2,
\end{equation}
above which $(N_f>N_f^c)$ the chiral symmetry restores
\footnote{In the quenched limit $N_f \rightarrow 0$, the theory has
only one phase, chiral-symmetry-breaking phase, as shown by
the SD equation \cite{KN90} and the Monte Carlo simulation
\cite{DKK89a}.}
and for $N_f<N_f^c$ the fermion mass is dynamically generated
and obeys the scaling of the essential singularity type:
\begin{equation}
 m \sim \Lambda f(N_f), \
f(N_f) = \exp \left[ - {2n\pi \over \sqrt{N_f^c/N_f-1}} \right] .
\end{equation}
This result was obtained under the assumption of no wavefunction
renormalization $A(p) \equiv 1$ for the fermion in the Landau gauge
where the fermion propagator is written as
$S(p)=[A(p) \rlap{$p$}/ -B(p)]^{-1}$.
They justified their approximations based on the argument of
$1/N_f$ expansion in which $A(p) = 1 + {\cal O}(1/N_f)$
and $\Gamma_\mu(p,q)=\gamma_\mu + {\cal O}(1/N_f)$.
\par
However, this was criticised by Pennington and Webb (PW)
\cite{PW88} and  Atkinson, Johnson and Pennington (AJP)
\cite{AJP88}. They claimed that the one loop correction to the
wavefunction renormalization and the corresponding vertex correction
of the type $\Gamma_\mu(p,q) = \gamma_\mu A(q)$
\footnote{Similar ansatz for the vertex, e.g.,
$\Gamma_\mu(p,q) = {1 \over 2}\gamma_\mu [A(p)+A(q)]$ or
$\Gamma_\mu(p,q) =
\gamma_\mu
[A(p)\theta(p^2-q^2)+A(q)\theta(q^2-p^2)]$,  leads
to the same result.}
 leads to the scaling law of the exponential type:
\begin{equation}
 f(N_f) = \exp \left[ - {3\pi^2 \over 8} N_f  \right]
\end{equation}
and that there does not exist a finite critical flavour,
$N_f^c =\infty$.
Similar claim was stated
\cite{Matsuki91} in the approach of the effective potential
in contrast to the original result by Matsuki, Miao and
Viswanathan \cite{MMV88}.
\par
On the other hand, the Monte Carlo simulation of non-compact QED on
a lattice by Daggoto, Kogut and Kocic \cite{DKK89a}  suggests
the existence of two phases which are separated by a finite $N_f^c
\sim 3.5 \pm 0.5$.  However Azcoiti et al. \cite{Azcoiti93} recently
report different and delicate results.  Hence this issue has not yet
been confirmed by the simulation of lattice non-compact QED.
\par
In the framework of the SD equation, it was also pointed out that
the inclusion of the infrared (IR) cutoff which plays the role of
the finite size of the lattice simulation may drastically change the
critical phenomenon associated with the dynamical symmetry
breaking.  Especially the ansatz of PW leads to the
mean-field type scaling law and the cutoff-dependent finite
$N_f^c$ in the large IR cutoff,  while the exponential type
scaling and infinite $N_f^c$ are obtained in the
sufficiently small IR cutoff \cite{KN92}.
\par
The subtlety of the problem in the framework of the SD equation
comes from the fact that, in order to take into account the
wavefunction renormalization,  it is indispensable to include the
vertex correction as a consequence of the Ward-Takahashi (WT)
identity.  However incorporating the vertex correction properly is
quite difficult in the non-perturbative study of gauge theory,
although there are many works of going along such a line
 under the appropriate ansatz for the vertex
in QED3 \cite{CPW92,AJM90} as well as the (3+1)-dimensional case,
QED4 \cite{CP91,Kondo92,ABGPR93}.

\par
In this paper we investigate this problem from a different point of
view.  Instead of considering the vertex correction, we look for the
situation in which we do not have to take into account
the vertex correction.
This is achieved only when there is no wavefunction
renormalization.  In the quenched approximation which neglects the
vacuum polarization in the photon propagator, this is simply
realized by taking the Landau gauge \cite{KN89}.  In the presence of
the vacuum polarization and/or the Chern-Simons (CS) term
\cite{DJT81}, however, we cannot  choose the (covariant) gauge
fixing parameter such that
$A(p) \equiv 1$ follows. Recently it has been recognized that this
can be done by taking the {\it non-local} gauge
\cite{GSC90,Simmons90,KM92}.  This standpoint leads to the
complicated non-local gauge instead of the vertex correction.
However this method has some advantages that we do not have to
treat the coupled SD equation for two functions $A(p)$ and
$B(p)$, and somewhat arbitrary (non-perturbative) ansatz for the
vertex so as to satisfy the WT identity.
\par
In the same scheme, we study the dynamical breakdown of the parity.    The bare
CS term breaks the parity explicitly. Even if there exists no bare CS term, the
CS term may be induced by the radiative correction through the
non-perturbative effect.  Therefore there is a possibility that the
induced CS term breaks the parity. However previous studies so far
support the claim that the  dynamical breakdown of the parity does
not occur  \cite{ABKW86b,Poly88,RY86a,HM89}
in agreement with the general argument \cite{VW84}.   We clarify
this reason from the viewpoint of SD equation.


\section{SD equation and the non-local gauge}
We consider the (2+1)-dimensional QED (QED3) with the following
(euclidean) lagrangian
\begin{eqnarray}
{\cal L}_{QED3} = \bar \psi^i (i\gamma_\mu \partial_\mu
- m_e - m_o \tau) \psi^i + {1 \over 4} F_{\mu \nu}^2
+  e \bar \psi^i \gamma_\mu \psi^i A_\mu
+ {\cal L}_{CS} + {\cal L}_{GF},
\end{eqnarray}
with the Chern-Simons term
\begin{eqnarray}
{\cal L}_{CS}
= {i \over 2}
\theta \epsilon_{\mu \nu \rho} A_\mu \partial_\nu A_\rho,
\end{eqnarray}
and the gauge fixing term ${\cal L}_{GF}$ whose form will be
specified in what follows.
According to Appelquist et al. \cite{ABKW86a},  $\psi^i $
denotes the 4-component Dirac fermion with a flavour index
$i(i=1,...,N_f)$.  Here
$\gamma_a (a=0,1,2)$ are $4 \times 4$ matrices which
satisfy the clifford algebra
$\{\gamma_\mu, \gamma_\nu \} = -2 \delta_{\mu\nu}$
and  are explicitly defined by
\footnote{In what follows, $f:=g$ implies that $f$ is
defined by $g$.}
\begin{eqnarray}
 \gamma_0 := \left( \matrix{ -i \sigma^3 & 0           \cr
                            0          & i \sigma^3 \cr }
\right),
\gamma_1 := \left( \matrix{ i \sigma^1 & 0           \cr
                            0          & -i \sigma^1 \cr } \right),
\gamma_2 := \left( \matrix{ i \sigma^2 & 0           \cr
                            0          & -i \sigma^2 \cr } \right),
\end{eqnarray}
with $\sigma_a (a=1,2,3)$ being the Pauli matrices.
We further introduce the $4 \times 4$ matrices $\gamma_3$,
$\gamma_5$ and $\tau$ as
\footnote{Note that $\tau$ anticommutes with $\gamma_3, \gamma_5$
and commutes with $\gamma_0, \gamma_1, \gamma_2$, and that
$\gamma_5^\dagger = - \gamma_5$, $\gamma_5^2 = -1$.}
\begin{eqnarray}
 \gamma_3 := \left( \matrix{ 0 & 1 \cr
                            1 & 0 \cr } \right),
 \gamma_5 :=  \gamma_0 \gamma_1 \gamma_ 2 \gamma_3
 = \left( \matrix{ 0 & -1 \cr
                   1 &  0 \cr } \right),
 \tau := \gamma_3 \gamma_5 = \left( \matrix{ 1 &  0 \cr
                                             0 & -1 \cr } \right).
\end{eqnarray}

\par
First we define the chiral transformation by
$ \psi(x) \rightarrow \gamma_3 \psi(x),
 A_\mu(x) \rightarrow  A_\mu(x)  .
$
Then the mass term $m_e \bar \psi \psi$ and $m_o \bar \psi \tau
\psi$ are respectively odd and even under the chiral transformation.
Next the parity transformation is defined by
$
 \psi(x_0,x_1,x_2) \rightarrow P \psi(x_0,-x_1,x_2),
 A_\mu(x_0,x_1,x_2)
\rightarrow  (-1)^{\delta_{\mu 1}} A_\mu(x_0,-x_1,x_2),
$
with $P=-i\gamma_5 \gamma_1$.
Then $m_e \bar \psi \psi$ and $m_o \bar \psi \tau \psi$ are
respectively parity even and odd mass term.
The bare CS term ${\cal L}_{CS}$ is parity odd and hence the
inclusion of the bare CS term breaks the parity explicitly even if
$m_o=0$.
\par
Corresponding to the diagram of Fig.1, the SD equation for the
full (exact) fermion propagator $S(p) = [A(p) \rlap{$p$}/ - B(p)]^{-1}$
is written as follows.
\begin{eqnarray}
  A(p) \rlap{$p$}/ - B(p) &=& \rlap{$p$}/ - m_e - m_o \tau
\nonumber\\
&& + e^2 \int {d^3q \over (2\pi)^3} D_{\mu \nu}(q-p)
\gamma_\mu {A(q) \rlap{$q$}/ - B(q) \over A^2(q) q^2 + B^2(q)}
\Gamma_\nu(p,q) ,
\end{eqnarray}
where $D_{\mu \nu}(k)$ is the full photon propagator and
$\Gamma_\nu(p,q)$ the full vertex function. In this paper we take the
bare vertex $\gamma_\mu$ instead of the full vertex
$\Gamma_\mu(p,q)$. In order for the bare vertex approximation
$\Gamma_\mu(p,q)=\gamma_\mu$
to be justified in this SD equation, we must guarantee no
wavefunction renormalization
$A(p) \equiv 1$ in light of the Ward-Takahashi (WT) identity:
$(p_\mu - q_\mu) \Gamma_\mu(p,q) = S^{-1}(p)-S^{-1}(q)$.
Actually, in the case of the quenched approximation (without CS
term):
\begin{equation}
D_{\mu \nu}(k) = D_{\mu \nu}^0(k) \equiv
{1 \over k^2} \left[ \delta_{\mu \nu}-{k_\mu k_\nu \over
k^2} (1-\xi) \right]
\end{equation}
for the covariant gauge-fixing
${\cal L}_{GF}={1 \over 2\xi} (\partial_\mu A_\mu)^2$,
this is simply achieved by taking the Landau gauge $\xi=0$, see e.g.
\cite{KN90}.
\par
In the presence of the vacuum polarization and/or the
CS term, this cannot be achieved by choosing an appropriate value for
the usual gauge fixing parameter $\xi$ and instead we must adopt the
non-local gauge in the sense that $\xi$ becomes a function of the
photon momentum: $\xi=\xi(k^2)$ \cite{GSC90,Simmons90,KM92}.
In QED3,
therefore, we consider the photon propagator of the form:
\begin{eqnarray}
D_{\mu \nu}(k) =   \left[
\delta_{\mu \nu}-{k_\mu k_\nu \over k^2} \right] D_T(k)
+ {k_\mu k_\nu \over k^2} D_L(k)
+ \epsilon_{\mu \nu \rho} {k_\rho \over \sqrt{k^2}} D_O(k),
\end{eqnarray}
with the non-local gauge
\footnote{This non-local gauge function $\xi$ reduces to
the usual gauge fixing parameter $a$ of the covariant
gauge
${1 \over 2a}(\partial_\mu A_\mu)^2$ in the limit
$D_T(k) \rightarrow {1 \over k^2}$ and
$D_L(k) \rightarrow {a \over k^2}$.}
\begin{eqnarray}
\xi(k^2) =  {D_L(k) \over D_T(k)}.
\end{eqnarray}
\par
The vacuum polarization tensor in QED3 is
written  as
\begin{eqnarray}
\Pi_{\mu \nu}(k) =   \left[
\delta_{\mu \nu}-{k_\mu k_\nu \over k^2} \right] \Pi_T(k)
+ {k_\mu k_\nu \over k^2} \Pi_L(k)
+ \epsilon_{\mu \nu \rho} {k_\rho \over \sqrt{k^2}} \Pi_O(k).
\end{eqnarray}
 From the SD equation for the photon propagator,
\begin{equation}
[D_{\mu \nu}(k)]^{-1}=[D_{\mu \nu}^0(k)]^{-1}-\Pi_{\mu
\nu}(k),
\end{equation}
the general form for $D_T(k)$ and $D_O(k)$ is
written as
\begin{eqnarray}
 D_T(k) =  {k^2-\Pi_T(k) \over
(k^2-\Pi_T(k))^2+(\Pi_O(k)-\theta \sqrt{k^2})^2},
\end{eqnarray}
\begin{eqnarray}
 D_O(k) =  {\Pi_O(k) -\theta \sqrt{k^2} \over
(k^2-\Pi_T(k))^2+(\Pi_O(k)-\theta \sqrt{k^2})^2}.
\end{eqnarray}
\par
 Now we define the parity eigenmatrix decomposition by
$\chi_{\pm} = {1 \over 2}(1\pm \tau)$.  Then we have
$A \rlap{$p$}/ - B
= A_e \rlap{$p$}/ + A_o \tau \rlap{$p$}/
- B_e - B_o \tau
= (A_{+} \rlap{$p$}/ - B_{+}) \chi_{+}
+ (A_{-} \rlap{$p$}/ - B_{-}) \chi_{-}$ and
$[A \rlap{$p$}/ - B]^{-1}=
[A_{+} \rlap{$p$}/ - B_{+}]^{-1} \chi_{+}
+ [A_{-} \rlap{$p$}/ - B_{-}]^{-1} \chi_{-}$.
Then the parity even and odd parts are obtained as
$A_e=(A_{+}+A_{-})/2$ and $A_o=(A_{+}-A_{-})/2$ and similarly for
$B_e$ and $B_o$: $B_e=(B_{+}+B_{-})/2$ and $B_o=(B_{+}-B_{-})/2$.
 Thus the SD equation for the fermion propagator is
decomposed into the simultaneous integral equation for $A$ and $B$:
\begin{eqnarray}
 A_{\pm}(p) = 1 &+& {e^2 \over p^2} \int {d^3 q \over (2\pi)^3}
{1 \over q^2 A_{\pm}^2(q)+B_{\pm}^2(q)}
\Biggr\{ \pm 2 B_{\pm}(q) {p\cdot (q-p) \over |q-p|} D_O(k)
\nonumber\\ &&
+ A_{\pm}(q) \left[ (p \cdot q) \xi(k^2) + 2{(k\cdot q)(k \cdot p)
\over k^2} (1-\xi(k^2)) \right] D_T(k)  \Biggr\} ,
\end{eqnarray}
\begin{eqnarray}
 B_{\pm}(p) = m_{\pm} &+&  e^2  \int {d^3 q \over (2\pi)^3}
{1 \over q^2 A_{\pm}^2(q)+B_{\pm}^2(q)}
\nonumber\\ &&
\left\{ B_{\pm}(q) [2+\xi(k^2)] D_T(k)
\mp 2 A_{\pm}(q) {q\cdot (q-p) \over |q-p|} D_O(k) \right\},
\label{SDB}
\end{eqnarray}
where
 $m_{\pm} := m_e \pm m_o$ and $k:=q-p$.
\par
As shown in Appendix A, if we take the following
non-local gauge function $\xi(s:=k^2)$:
\begin{eqnarray}
 \xi(s) = 2 -
{2 \over s^2 D_T(s)} \int_0^s dt  D_T(\sqrt{t})t ,
\label{nonlocalf}
\end{eqnarray}
then we have
\begin{eqnarray}
 A_{\pm}(p) = 1  \pm {e^2 \over p^2} \int {d^3 q \over
(2\pi)^3} {2 B_{\pm}(q) \over q^2
A_{\pm}^2(q)+B_{\pm}^2(q)} D_O(k)
  {p\cdot (q-p) \over |q-p|}  .
\label{SDA0}
\end{eqnarray}
 If we do not take into account the CS term $D_O(k) \equiv
0$, the above expression for the non-local gauge gives
exactly $A_{\pm} \equiv 1$ for any $B$.
  In the case of the induced CS term, i.e., $\theta=0$ (see
section 4), the term  $B_{\pm}(p) D_O(k)$ is of the order
$B^2$, which is negligible in the linearized SD equation
which we are going to study in what follows.
Even in the presence of bare CS term (see section 3),
the contribution from the second term in the right hand
side of eq.~(\ref{SDA0}) is very small compared with 1
 in the neighborhood of the critical point of the chiral phase
transition which we pay special attention to in this paper.
 \footnote{This has been confirmed numerically, see
\cite{KM94}.}

\par
If we adopt this non-local gauge, therefore, we obtain
$
 A_{e}(p) \cong 1
$
and
$
 A_{o}(p) \cong 0,
$
and the even and odd mass functions obey
\begin{eqnarray}
 B_{e}(p) &=& m_{e} +  e^2  \int {d^3 q \over (2\pi)^3} {1
\over [q^2+B_{e}^2(q)+B_{o}^2(q)]^2-4B_{e}^2(q)B_{o}^2(q)}
\nonumber\\ && \times
\Biggr\{ [2+\xi(k^2)] [q^2 +B_{e}^2(q)-B_{o}^2(q)]B_{e}(q)
D_T(k)
\nonumber\\ &&  - 4B_{e}(q)B_{o}(q) {q\cdot (q-p) \over
|q-p|} D_O(k) \Biggr\},
\label{SDBe}
\end{eqnarray}
\begin{eqnarray}
 B_{o}(p) &=& m_{o} +  e^2  \int {d^3 q \over (2\pi)^3} {1
\over [q^2 +B_{e}^2(q)+B_{o}^2(q)]^2-4B_{e}^2(q)B_{o}^2(q)}
\nonumber\\ && \times
\Biggr\{ [2+\xi(k^2)] [q^2 -B_{e}^2(q)+B_{o}^2(q)]B_{o}(q)
D_T(k)
\nonumber\\ && + 2[q^2 +B_{e}^2(q)+B_{o}^2(q)]{q\cdot (q-p)
\over |q-p|} D_O(k) \Biggr\},
\label{SDBo}
\end{eqnarray}

\par
In the non-local gauge given above, $B_{\pm}(p)$ obey
\begin{eqnarray}
 B_{\pm}(p) = m_{\pm} + {e^2 \over 2\pi^2} \int_0^\Lambda
 dq {q^2 \over q^2+B_{\pm}^2(q)} \left[ B_{\pm}(q) K(p,q)
 \mp L(p,q) \right],
\label{SD+-}
\end{eqnarray}
where we have introduced the ultraviolet (UV) cutoff
\footnote{It is shown that the solution damps very
quickly in the region
$p>\alpha$ and hence the UV cutoff can be
substantially identified with $\alpha$
\cite{ABKW86a,ANW88}.}
$\Lambda$ and defined
\begin{eqnarray}
 K(p,q) := \int_{0}^{\pi} d \vartheta \sin \vartheta
 \left[ 1+{\xi(q-p) \over 2} \right] D_T(q-p),
\label{kernelK}
\end{eqnarray}
and
\begin{eqnarray}
 L(p,q) :=  \int_{0}^{\pi} d \vartheta \sin \vartheta
 {q \cdot (q-p) \over |q-p|} D_O(q-p),
\label{kernelL}
\end{eqnarray}
for the angle $\vartheta$ defined by
$(q-p)^2=q^2+p^2-2pq \cos \vartheta$.
\par
 This integral equation is non-linear in
$B_{\pm}$ and  difficult to be  solved analytically.
Therefore we adopt the approximation to get the appropriate
linear equation.
\footnote{In the linearization of eq.~(\ref{SDBe}) we
should keep the coupling between $B_e(q)$ and $B_o(q)$,
which leads to an infrared cutoff depending on both
$B_e(0)$ and $B_o(0)$.
In the linearization of eq.~(\ref{SD+-}), leading to
eq.~(\ref{linearSD0}), the equations for $B_{+}$ and
$B_{-}$ decouple completely, but it should be noted that
$M_{+}$ and  $M_{-}$ are different from each other due to
the presence of
$\theta$ \cite{Maris94}.}
Here it should be remarked that, in the linearized
equation, if
$B(p)$ is a solution, the rescaled function $\kappa B(p)$ by
a constant $\kappa$ is also a solution when
$L(p,q) \equiv 0$ and $m_{\pm}=0$.  Hence the magnitude of
the linearized integral equation can not be determined in
this case. Therefore, in the linearized equation, we must
introduce an additional condition, the normalization
condition, which specifies the magnitude of the solution.
\par
According to \cite{Atkinson,Kondo92p}, we linearize
eq.~(\ref{SD+-}) by replacing  the denominator $q^2+B^2(q)$
with $q^2$ and introduce the infrared (IR) cutoff $M$ in
the region of integration: $M \le q \le \Lambda$.
Then we adopt the normalization condition
$M=B(p=M)$.  This procedure enforces us to consider the
solution only in the region: $M \le p \le \Lambda$.
In this approximation, the SD equation reduces to
\begin{eqnarray}
 B_{\pm}(p) = m_{\pm} + {e^2 \over 2\pi^2}
 \int_{M_{\pm}}^\Lambda
 dq \left[ B_{\pm}(q) K(p,q) \mp L(p,q) \right],
\label{linearSD0}
\end{eqnarray}
supplemented with the normalization
condition $M_{\pm}=B_{\pm}(p=M_{\pm})$.
\par


\section{Chiral symmetry breaking}
First of all, we consider the dynamical breaking of the chiral
symmetry.  In this section, we neglect the induced CS term,
$\Pi_O(k) \equiv 0$ and take into account the one-loop vacuum
polarization:
\begin{eqnarray}
 \Pi_T(k) = - \alpha \sqrt{k^2} + {\cal O}(B^2),
\ \alpha := {e^2 N_f \over 8}.
\end{eqnarray}
  Then $D_T(k)$ and $D_O(k)$ read
\begin{eqnarray}
D_T(k)  =  {k^2+\alpha \sqrt{k^2} \over
(k^2+\alpha \sqrt{k^2})^2+\theta^2 k^2},
D_O(k)  =  - {\theta \sqrt{k^2} \over
(k^2+\alpha \sqrt{k^2})^2+\theta^2 k^2},
\end{eqnarray}
where $k^2=(p-q)^2:=s$ denotes the squared photon momentum.
 In this case, eq.~(\ref{nonlocalf}) leads to the non-local gauge
function:
\begin{eqnarray}
 \xi(s) &=& 2 - 2 {(s+\alpha \sqrt{s})^2+\theta^2 s \over
s^2(s+\alpha \sqrt{s})} \Biggr\{ -2 \alpha \sqrt{s} + s
- 4 \alpha \theta \left[ \arctan {\alpha \over \theta}
- \arctan {\alpha + \sqrt{s} \over \theta} \right]
\nonumber\\
&& + (\alpha^2-\theta^2) \ln {(\alpha+\sqrt{s})^2+\theta^2 \over
\alpha^2+\theta^2} \Biggr\} .
\label{nonlocalg}
\end{eqnarray}
The function $\xi=\xi(k^2)$ of $k^2$ is not singular even at $k^2=0$
and decreases to zero as $k^2 \rightarrow \infty$, see Fig.2.
The non-local gauge just obtained eq.~(\ref{nonlocalg})
has quite complicated form.
In order to solve the SD equation analytically, therefore, we expand
the  non-local gauge eq.~(\ref{nonlocalg}) around $k^2=0$ and
take into account the first and the second terms, since it changes
rapidly in the infrared (IR) region $k/\Lambda \ll 1$.  The effect
of higher order terms will be discussed in the final section.  Of
course, it is necessary to perform the numerical calculation to
include the non-local gauge completely, which is a subject of the
subsequent paper \cite{KM94}. It is easy to show that
$\xi(k^2)$ of  eq.~(\ref{nonlocalg}) has the
following expansion in the IR region $k \ll \alpha$:
\begin{eqnarray}
 \xi(k^2) = {2 \over 3} - {1 \over 3} {\alpha^2-\theta^2 \over
\alpha^2+\theta^2}{\sqrt{k^2} \over \alpha}
+ {\cal O}\left( \left({k \over \alpha} \right)^2 \right).
\label{nonlocalexpa}
\end{eqnarray}
This expansion can be matched with the $1/N_f$ expansion  (when
$\theta=0$).  If we expand the non-local gauge into such a series and
truncate the series up to  ${\cal O}((k/\alpha)^N), (N \ge 1)$, the
truncated non-local gauge $\xi$ guarantees that $A(p) = 1 + {\cal
O}((1/N_f)^{N+1})$.  This implies that the WT identity is satisfied at
least up to ${\cal O}((1/N_f)^{N+1})$ by the vertex function
$\Gamma_\mu(p,q) = \gamma_\mu + {\cal O}((1/N_f)^{N+1})$ and the
wavefunction renormalization function $A(p) = 1 + {\cal
O}((1/N_f)^{N+1})$. The first term $2/3$ corresponds to the gauge choice
of Carena, Clark and Wagner \cite{CCW91}, which is obtained as a
consequence of our systematic treatment of non-local gauge.
\par
 We substitute the above form eq.~(\ref{nonlocalexpa}) of
the non-local gauge into the SD equation
eq.~(\ref{SDB}) for $B_{\pm}$.
In order to solve the equation analytically, we linearize
the equation based on the second approximation explained in the
previous section:
\begin{eqnarray}
 B_{\pm}(p) = m_{\pm} \pm  f(p)
+ {4\alpha \over \pi^2 N_f}
 \int_{M_{\pm}}^\Lambda
 dq  B_{\pm}(q) K(p,q) ,
\end{eqnarray}
with
\begin{eqnarray}
 f(p) :=  -{4\alpha \over \pi^2 N_f}
 \int_{M_{\pm}}^\Lambda dq  L(p,q) .
\end{eqnarray}
Here $K(p,q)$ is obtained by using eq.~(\ref{nonlocalexpa})
as
\begin{eqnarray}
 K(p, q) = {8 \over 3} {\alpha \over \alpha^2+\theta^2}
\left[ {p+q-|p-q| \over 2pq}  - {9 \over 8}
{\alpha^2-\theta^2 \over \alpha^2+\theta^2} {1 \over
\alpha} \right],
\label{kernelK2}
\end{eqnarray}
where, corresponding to eq.~(\ref{nonlocalexpa}) of the
non-local gauge-function,  we have used the following
expansion:
\begin{eqnarray}
 D_T(k) = {\alpha \over \alpha^2+\theta^2}{1 \over
\sqrt{k^2}}  - {\alpha^2-\theta^2 \over
(\alpha^2+\theta^2)^2} + {\cal O}(\sqrt{k^2}) .
\end{eqnarray}
Similarly $L(p,q)$ is obtained as
\begin{eqnarray}
 L(p, q) &=& -{\theta \over \alpha^2+\theta^2}
\left[ 1 - {p^2-q^2 \over 2pq} \ln {p+q \over |p-q|}
\right]
\nonumber\\&&
+ {2\alpha \theta \over  (\alpha^2+\theta^2)^2}
\Biggr[ {q \over p}(p+q-|p-q|)
\nonumber\\&&
 -{(p+q)(p^2-pq+q^2)-|p-q|(p^2+pq+q^2) \over 3pq} \Biggr],
\end{eqnarray}
where we have used
\begin{eqnarray}
 D_O(k) = -{\theta \over \alpha^2+\theta^2}{1 \over
\sqrt{k^2}}  + {2\alpha \theta \over
(\alpha^2+\theta^2)^2} + {\cal O}(\sqrt{k^2}).
\end{eqnarray}

Thus we obtain the SD equation
\begin{eqnarray}
 B_{\pm}(p) = m_{\pm} \pm f(p) + \lambda
 \int_{M_{\pm}}^\Lambda dq
 B_{\pm}(q) \left[ {p+q-|p-q| \over 2pq}  - {9 \over 8}
{\alpha^2-\theta^2 \over \alpha^2+\theta^2} {1 \over
\alpha} \right],
\label{linearSD}
\end{eqnarray}
with
\begin{eqnarray}
 \lambda := {32 \over 3\pi^2N_f}
{\alpha^2 \over \alpha^2+\theta^2} .
\end{eqnarray}
Note that, if $\theta=0$, $f(p) \equiv 0$ and
$B_{\pm}(p)$ obeys the same equation.
\par
The integral equation eq.~(\ref{linearSD}) can be solved as
the boundary value problem of the second order differential
equation as follows.
 It can be converted into the differential equation
\begin{eqnarray}
 {d \over dp} \left( p^2 {d B_{\pm}(p) \over dp} \right)
 + \lambda B_{\pm}(p)
= \pm r(p) ,
\label{de}
\end{eqnarray}
where
\begin{eqnarray}
 r(p) := {d \over dp} \left( p^2 {d f(p) \over dp} \right).
\end{eqnarray}
Then the solution of the SD integral
equation can be obtained from the general solution of the
differential equation eq.~(\ref{de}) which satisfies both
the IR boundary condition (BC):
\begin{eqnarray}
 B_{\pm}'(M_{\pm}) = \pm f'(M_{\pm}),
\label{IRBC}
\end{eqnarray}
and the UV BC:
\begin{eqnarray}
 m_{\pm} = B_{\pm}(\Lambda) + \left[ 1 - J(\alpha,\theta)
\right] \Lambda [B_{\pm}'(\Lambda) \mp f'(\Lambda)]
\mp f(\Lambda),
\label{UVBC}
\end{eqnarray}
where
\begin{eqnarray}
 J(\alpha, \theta) :=  {9 \over 8}
{\alpha^2-\theta^2 \over
\alpha^2+\theta^2} {\Lambda \over \alpha}.
\label{defJ}
\end{eqnarray}

\par
Here it should be remarked that the second term in the
non-local gauge function  eq.~(\ref{nonlocalexpa}) does
not change the differential equation eq.~(\ref{de}) and
simply changes the UV BC.  Therefore the second term takes
no effect on the general solution of the differential
equation, but may change the scaling law through the UV BC.
Without it, UV BC reads
$ m_{\pm} = B_{\pm}(\Lambda) +  \Lambda
[B_{\pm}'(\Lambda) \mp f'(\Lambda)] \mp f(\Lambda)$.
\par
If $r(p) \equiv 0$, the general solution of eq.~(\ref{de})
is given by the linear combination of two special
solutions:
$p^{(-1+\tau)/2}$ and $p^{(-1-\tau)/2}$ where
$\tau:=\sqrt{1-4\lambda}$.
In the presence of the inhomogeneous term $r(p)$,
the general solution of eq.~(\ref{de}) is given by
\begin{equation}
 B_{\pm}(p) = M_{\pm} \left[
c_1 \left({p \over M_{\pm}}\right)^{{-1+\tau \over 2}} +
c_2 \left({p \over M_{\pm}}\right)^{{-1-\tau \over 2}}
\right]
\pm b_\theta(p),
\label{solution}
\end{equation}
where
\begin{equation}
b_\theta(p) :=  p^{{-1+\tau \over 2}} \int^p dq \
q^{-1-\tau}
\int^q dk  r(k) k^{{-1+\tau \over 2}}.
\end{equation}
Two constants $c_1$ and $c_2$ are determined from the IR BC
and the normalization condition in what follows.
\par
In order to write $b_\theta(p)$ explicitly, we must
calculate the integration as
\begin{equation}
 f(p)  =   {3\lambda \theta \over 8\alpha}
\int_0^\Lambda dq
 \left[ 1 - {p^2-q^2 \over 2pq} \ln {p+q \over |p-q|}
\right].
\label{functionf}
\end{equation}
 Numerical integration shows that the function
$f(p)$ is well approximated in the interval
$[0,\Lambda]$ by the function:
\begin{equation}
f(p)  =   {3\lambda \theta \over 8\alpha} \Lambda
 \left[ 2 - a {p \over \Lambda}
+ b  ({p \over \Lambda})^2 \right],
\end{equation}
with $a=2.5329$ and $b=0.7913$, see Fig.~3.
Especially, in the region $p \ll \Lambda$, $f(p)$ is well
approximated by the linear function with $a=2, b=0$. This
can be derived by expanding the integrand of $f(p)$ in
powers of $q/p$ or $p/q$ after dividing the integral
region into $p>q$ and $p<q$, respectively.
\par
Hence the function $r(p)$ reads
\begin{equation}
r(p)  =   {3\lambda \theta \over 4\alpha} \Lambda
 \left[ - a {p \over \Lambda}
+ 3b  \left({p \over \Lambda}\right)^2
\right].
\end{equation}
This leads to
\begin{equation}
b_\theta(p)  =  {3\lambda \theta \over 4\alpha}\Lambda
 \left[ - {a \over 2+\lambda}{p \over \Lambda}
+ {3b \over 6+\lambda}
\left({p \over \Lambda}\right)^2 \right] .
\label{btheta}
\end{equation}
Note that $b_\theta(p)$ is monotonically decreasing in $p$
and $b_\theta(p) \downarrow 0$ as $\theta \downarrow 0$
uniformly in $p$.

\par
By using $f'(M)={3\lambda \theta \over 4\alpha}
\left(-{a \over 2}+b{M \over \Lambda} \right)$,
IR BC eq.~(\ref{IRBC})  reads
\begin{equation}
   c_1 {-1+\tau \over 2}+c_2 {-1-\tau \over 2}
= \pm  {3\lambda \theta \over 4\alpha}
\left[ - {\lambda \over 2(2+\lambda)}
+ b{\lambda \over 6+\lambda} {M_{\pm} \over \Lambda}
\right].
\end{equation}
{}From the normalization condition,
$B_{\pm}(M_{\pm})=M_{\pm}$, we obtain
\begin{equation}
   c_1+c_2  = 1 \pm
{3\lambda \theta \over 4\alpha}
\left[  a{1 \over 2+\lambda} - b{3 \over
6+\lambda} {M_{\pm} \over \Lambda} \right].
\end{equation}
Therefore two coefficients $c_1, c_2$ are determined as
\begin{eqnarray}
 c_1 &=&
{1+\tau \over 2\tau} \pm {3\lambda \theta \over
4\alpha\tau}
\left[  a{1-\lambda+\tau \over 2(2+\lambda)} + b{M_{\pm}
\over \Lambda}{-3+2\lambda-3\tau \over 2(6+\lambda)}
\right],
\nonumber\\
 c_2 &=&  {-1+\tau \over 2\tau} \pm {3\lambda \theta \over
4\alpha\tau}
\left[  a{-1+\lambda+\tau \over 2(2+\lambda)} + b{M_{\pm}
\over \Lambda}{3-2\lambda-3\tau \over 2(6+\lambda)}
\right] .
\label{coeff}
\end{eqnarray}

Finally the UV BC eq.~(\ref{UVBC}) is written as
\begin{eqnarray}
 {m_{\pm} \over \Lambda}
&=& [1+J+\tau(1-J)] {c_1 \over 2}
\left({M_{\pm} \over \Lambda}\right)^{{3-\tau \over 2}}
+ [1+J-\tau(1-J)] {c_1 \over 2}
\left({M_{\pm} \over \Lambda}\right)^{{3+\tau \over 2}}
\nonumber\\&&
\mp {3\lambda \theta \over 4\alpha}
C(\lambda, \theta),
\label{UVBC2}
\end{eqnarray}
where
\begin{eqnarray}
 C(\lambda, \theta) :=
1 - a{\lambda(2-J) \over 2(2+\lambda)}  +
b{\lambda(3-2J) \over 2(6+\lambda)}  .
\end{eqnarray}

\subsection{$\theta=0$}
In the absence of $\theta$, $B_{+}$ and
$B_{-}$ obey the same SD equation when $m_{\pm}=0$.
Therefore
 $B_e(p) \equiv B_{\pm}(p)$ and
$B_o(p) \equiv 0$.
In this case $B_{\pm}(p) \equiv 0$ is a trivial solution
of eq.~(\ref{SDB}).  Here we define $M_e:=M_{\pm}$.
Then the UV BC eq.~(\ref{UVBC2}) yields
\begin{eqnarray}
0 = [1+J+\tau(1-J)] {c_1 \over 2}
\left({M_{e} \over \Lambda}\right)^{{3-\tau \over 2}}   +
[1+J-\tau(1-J)] {c_2 \over 2}
\left({M_{e} \over \Lambda}\right)^{{3+\tau \over 2}} ,
\end{eqnarray}
when $m_{\pm}=0$.
By using the coefficient obtained in
eq.~(\ref{coeff}) :
\begin{eqnarray}
 c_1 =  {1+\tau \over 2\tau},
\ c_2 =  {-1+\tau \over 2\tau} ,
\end{eqnarray}
the ratio $M_e/\Lambda$ is obtained as
\begin{eqnarray}
\left( {M_e \over \Lambda} \right)^\tau
= {(1+\tau) \over (1-\tau)}
{[1+J+(1-J)\tau] \over [1+J-(1-J)\tau]}.
\end{eqnarray}
The right hand side (RHS) is greater than 1, since
$0<\tau:=\sqrt{1-4\lambda}<1$ and $J>0$.
Therefore this equation can not have the solution
$M_e/\Lambda<1$ for real $\tau$, i.e., $\lambda<1/4$.
\par
In the region $\lambda>\lambda_c:=1/4$, we define
\begin{eqnarray}
 \omega := \sqrt{4\lambda-1}, \tau = i\omega,
\ \lambda = {32 \over 3\pi^2 N_f}.
\end{eqnarray}
Then we look for the solution of the equation
\begin{eqnarray}
\left( {M_e \over \Lambda} \right)^{i\omega}
= {1+J-(1-J)\omega^2+2i\omega \over
1+J-(1-J)\omega^2-2i\omega}.
\end{eqnarray}
This equation has the solution $0<M_e/\Lambda<1$ of
\begin{eqnarray}
 {M_e \over \Lambda}
= \exp \left[ - {2n\pi - 2\phi \over \omega} \right],
\ \phi = \arctan {2\omega \over 1+J-(1-J)\omega^2}.
\end{eqnarray}
In this region the non-trivial solution is obtained from
eq.~(\ref{solution}) and eq.~(\ref{coeff}).
\par
Thus it is shown that, in the chiral limit $m_e=0$, the
non-trivial chiral-symmetry-breaking solution exists only
for $N_f<N_f^c$ as
\begin{eqnarray}
 B_e(p) = {M_e^{3/2} \over \sqrt{p}}  {\sin \left( {1 \over
2} \sqrt{N_f^c/N_f-1} \ln {p \over M_e} +
\delta \right) \over  \sin \left( \delta \right) },
\end{eqnarray}
with $\delta=\arctan \omega=\arctan \sqrt{N_f^c/N_f-1}$.
The scaling law for the fermion dynamical mass is given by
\begin{eqnarray}
 {M_e \over \Lambda}  = \exp \left[ - {2n\pi-2\phi \over
\sqrt{N_f^c/N_f-1}} \right],
\ \phi = \arctan {2\sqrt{N_f^c/N_f-1} \over 2- N_f^c/N_f
\left( 1 - {9 \over 8} {\Lambda \over \alpha} \right) },
\end{eqnarray}
 with a critical value for the flavour:
\begin{eqnarray}
 N_f^c = {128 \over 3\pi^2}.
\end{eqnarray}  This shows that there exists the finite
critical number of flavors
$N_f^c$ above which $(N_f>N_f^c)$ the chiral symmetry
restores.
\par
In the absence of CS term,  the value $N_f^c={128 \over
3\pi^2} \sim 4.3$ that we have just found   is 4/3 times as
large as the value $32/\pi^2$ of ANW \cite{ANW88} and
coincides with that of the $1/N_f$ analysis by Nash
\cite{Nash89}.
\footnote{Nash's non-local gauge corresponds to a special
case of ours:
$D_L(k)/D_T(k)=$ constant, independently of $k$.}
  This result clearly comes from the first term
$\xi=2/3$ of the non-local gauge eq.~(\ref{nonlocalexpa}).
 The second term of the non-local gauge does change neither
the critical value nor the type of the scaling law for the
fermion mass, the essential singularity at
$N_f^c$, and merely decreases the critical coefficient,
i.e., the magnitude of the fermion mass, since
$ M_e/\Lambda
 \sim K \exp \left[ - 2n\pi/\sqrt{N_f^c/N_f-1} \right] , \
K = \exp \left[ 4/\left(1 + {9 \over 8}
{\Lambda \over \alpha} \right) \right] ,
$ in the neighborhood of $N_f^c$.

\subsection{$\theta \not= 0$}
In this subsection we consider the region
$N_f<N_f^c={128 \over 3\pi^2}$.
By substituting eq.~(\ref{coeff}) into the UV BC
eq.~(\ref{UVBC2}), the equation of state is  obtained as
\begin{eqnarray}
 {m_{\pm} \over \Lambda}
&=& {({M_{\pm} \over \Lambda})^{{3-i\omega \over 2}} \over
4i\omega}
\left[  R_{\pm}(\lambda,\theta) + i I_{\pm}(\lambda,\theta)
\right]
 -{({M_{\pm} \over \Lambda})^{{3+i\omega \over 2}} \over
4i\omega}
\left[  R_{\pm}(\lambda,\theta) - i I_{\pm}(\lambda,\theta)
\right]
\nonumber\\&&
\mp  {3\lambda \theta \over 4\alpha}C(\lambda,\theta),
\label{eqofstate}
\end{eqnarray}
where we have defined
\begin{eqnarray}
 R_{\pm}(\lambda,\theta) &:=& 1+J-\omega^2(1-J)\pm
2\Theta
[(1-\lambda)(1+J)-\omega^2(1-J)],
\nonumber\\
I_{\pm}(\lambda,\theta) &:=& 2\omega \{ 1 \pm \Theta
[2-\lambda(1-J)] \},
\end{eqnarray}
with $J=J(\alpha, \theta)$ defined by eq.~(\ref{defJ}) and
\begin{eqnarray}
\Theta :=
{3a\lambda \over 8(2+\lambda)} {\theta \over \alpha}.
\end{eqnarray}
In deriving this equation eq.~(\ref{eqofstate}), we have
neglected the term with the factor $b$ in
eq.~(\ref{coeff}), since
$b(M/\Lambda)^2 \ll a(M/\Lambda)$.
\par
When $m_{\pm}=0$, we look for the solution for
\begin{eqnarray}
\zeta_{\pm} :=  {M_{\pm} \over \Lambda} .
\end{eqnarray}
{}From the equation of state, $\zeta_{\pm}$ obeys
\begin{eqnarray}
 \sin ( \phi_{\pm} - {\omega \over 2} \ln \zeta_{\pm})
= \pm  {3 \over 2}
C(\lambda,\theta) \omega K_{\pm}^{-1/2}
 {\lambda \theta \over \alpha} \zeta_{\pm}^{-3/2} ,
\label{alge}
\end{eqnarray}
where we have defined
\begin{eqnarray}
\phi_{\pm} := \arctan {I_{\pm}(\lambda,\theta)
\over R_{\pm}(\lambda,\theta)},
\end{eqnarray}
and
\begin{eqnarray}
 K_{\pm} = R_{\pm}(\lambda,\theta)^2
+ I_{\pm}(\lambda,\theta)^2.
\end{eqnarray}
For the non-trivial solution to exist, the above equation
must have the solution for $0<\zeta_{\pm}<1$.
\par
We start from the limit $\theta \rightarrow 0$.
In this limit, RHS of eq.~(\ref{alge}) vanishes. Hence the
solution of eq.~(\ref{alge}) is given by
\begin{eqnarray}
t_{\pm} := \phi_{\pm} - {\omega \over 2} \ln \zeta_{\pm}
= n\pi \ (n=1,2,...).
\end{eqnarray}
Then, in the region $N_f<N_f^c$ and $\theta=0$,
$M_{\pm}$ obeys the scaling law of the essential
singularity type:
\begin{eqnarray}
  M_{\pm}  = \Lambda \exp \left[ - {2n\pi - 2\phi_{\pm}
\over \omega} \right] > 0,
\end{eqnarray}
as shown in the previous subsection.
{}From the argument \cite{ANW88,Miransky85} using the
effective potential of Cornwall-Jackiw-Tomboulis
\cite{CJT74},
 it is shown that the solution with $n=1$ corresponds to
the nodeless solution and gives the ground state solution
and that the solution with $n \ge 2$ corresponds to the
excited solution with nodes $n-1$.  The magnitude of the
solution with nodes $n$ is related to the ground state
solution as
\begin{eqnarray}
  {M_{\pm} \over \Lambda}
  = e^{{2\phi_{\pm} \over \omega}} \left(
e^{-{2\phi_{\pm} \over \omega}} {M_{\pm}^{n=1} \over
\Lambda} \right)^n .
\label{hiera}
\end{eqnarray}
This implies the hierarchy of the mass scale:
 $M_{\pm}^{n=1}>M_{\pm}^{n=2}>M_{\pm}^{n=3}>...$.

\par
Next we keep the value of $N_f$ below the critical
point $N_f^c$ and increases $\theta>0$.
To find the solution, we rewrite eq.~(\ref{alge}) as
\begin{eqnarray}
 \sin t_{\pm}  = H_{\pm}(t_{\pm}),
\ H_{\pm}(t) := \pm  {3 \over 2}
C(\lambda,\theta) \omega K_{\pm}^{-1/2}
 {\lambda \theta \over \alpha}
e^{-3\phi_{\pm} \over \omega} \exp \left[{3t \over \omega}
\right] .
\label{alge2}
\end{eqnarray}
The solution of eq.~(\ref{alge}) is given as the intersection point of the
$\sin t$ curve with the curve $H_{\pm}(t)$, see Fig.4.
In the absence of the CS term, $\theta=0$,
eq.~(\ref{alge}) has (countably) infinite
number of solutions as shown above.
However, in the precense of the CS term, $\theta\not=0$,
the strongly oscillating solutions with many nodes   ($n
\gg 1$) disappear and eq.~(\ref{alge}) has finite number of
solutions, no matter how $\theta$ is small, since
$|H_{\pm}(t)|$ is a rapidly and monotonically increasing
function in $t$ and
$H_{\pm}(t) \rightarrow \pm \infty$ as
$t \rightarrow \infty$.
\par
For small $\theta >0$,
$H_{+}(t)>0$ and $H_{-}(t)<0$ for our choice:
$a=2.5$ and $J$ given by eq.~(\ref{defJ}).   Paying
attention to the region $0<t<2\pi$, we see in this case
that
$t_{+}$ (resp. $t_{-}$) has a solution at
$t=\pi-\delta_{+}(\theta)$ (resp.
$t=\pi+\delta_{-}(\theta)$) slightly smaller (resp.
larger) than $\pi$ where $\delta_{\pm}(\theta) \ge 0$
and  $\delta_{\pm}(\theta=0)=0$, see Fig.~4.
Hence we obtain the solution
\begin{eqnarray}
  M_{\pm}  = \Lambda e^{\pm 2 {\delta_{\pm}(\theta) \over
\omega}}
\exp \left[ - {2n\pi - 2\phi_{\pm} \over \omega} \right] .
\label{scaling}
\end{eqnarray}
Therefore the solution
$M_{+}^{\theta}$ gets larger value than $M_{+}^{\theta=0}$,
while
$M_{-}^{\theta}$ gets smaller one than  $M_{-}^{\theta=0}$.
By substituting the coefficients
eq.~(\ref{coeff}) into eq.~(\ref{solution}), the
non-trivial solution is given by
\begin{eqnarray}
 B_{\pm}(p) = {M_{\pm}^{3/2} \over \sqrt{p}}
\left[ {\sin \left( {\omega \over 2} \ln {p \over M_{\pm}}
+ \delta \right) \over \sin \delta }
\pm 2 \Theta {\sin \left( {\omega \over 2} \ln {p \over
M_{\pm}} + \epsilon \right) \over \sin \epsilon }
\right] \pm b_\theta(p),
\end{eqnarray}
where $b_\theta(p)$ is given by eq.~(\ref{btheta})
and $\delta=\arctan \omega$ and
 $\epsilon=\arctan{\omega \over 1-\lambda}$.
This solution satisfies the normalization condition
$M_{\pm} = B_{\pm}(M_{\pm})$ as
should does.

\par
As discovered in \cite{KM94}, in the presence of the
Chern-Simons term, a first order phase transition does
occur.  This phenomenon can be explained in our scheme as
follows.
Consider the situation that $\theta$ is increased
gradually with $N_f$ being kept.
We find that there is a certain value of $\theta$, say
$\theta_c(N_f)$, such that at $\theta=\theta_c(N_f)$ the
nodeless solution $B_{-}(p)$ (corresponding to $n=1$) does
disappear suddenly, since $H_{-}(t)$ does not intersect
with $\sin t$ any more in the interval
$0<t<\pi$ for sufficiently large $\theta$.
At $\theta=\theta_c$, the solution $B_{-}(p)$ cease to
exist.
 For $\theta >\theta_c(N_f)$, only the nodeless solution
$B_{+}(p)$ exists.
This discontinuity at $\theta=\theta_c(N_f)$ persists for
$B_e=(B_{+}+B_{-})/2$ and $B_o=(B_{+}-B_{-})/2$.
This discontinuity can be observed as the first order phase
transition in the direction of $\theta$.
By using $S(p)=[A \rlap{$p$}/ - B]^{-1}= [A_{+} \rlap{$p$}/
- B_{+}]^{-1} \chi_{+}+[A_{-} \rlap{$p$}/ -B_{-}]^{-1}
\chi_{-}$, the chiral order parameter
$\langle \bar \psi \psi \rangle$ is written as
\begin{eqnarray}
\langle \bar \psi \psi \rangle/N_f
= \int_0^\Lambda
{p^2 dp \over \pi^2}\left[
 {B_{+}(p) \over p^2+B_{+}(p)^2}
+  {B_{-}(p) \over p^2+B_{-}(p)^2} \right].
\end{eqnarray}
Therefore such a first order transition can be observed as
the discontinuity of the chiral order parameter.

\par
The critical line is obtained as follows.
Substituting eq.~(\ref{scaling}) back into
eq.~(\ref{alge}), we obtain
\begin{eqnarray}
 {\lambda \theta \over \alpha}
= (-1)^{n+1}   {2 \over 3}
 \sin ( \delta_{\pm}(\theta))
{K_{\pm}^{1/2} \over \omega}
C(\lambda,\theta)^{-1}
  e^{\pm 3 {\delta_{\pm}(\theta) \over \omega}}
\exp \left[ - {3n\pi - 3\phi_{\pm} \over \omega} \right].
\label{criticalline}
\end{eqnarray}
Up to ${\cal O}(\theta)$, the critical line in the phase
diagram $(N_f,\theta)$ is given by
\begin{eqnarray}
 {\lambda \theta \over \alpha} \sim
\exp \left[ - {3n\pi \over \omega} \right].
\label{criticalline2}
\end{eqnarray}
Above this critical line in the phase diagram, the chiral
symmetry restores.
This shows that the CS term has the
effect to decrease the critical number of flavors,
$N_f^c(\theta) < N_f^c(\theta=0)$.
However the chiral-symmetry restoring transition is the
first order phase transition.
\par
At typical points in the phase
diagram, $(\theta, N_f)$, which is drawn schematically
in Fig.5, we find
($M_e=(M_{+}+M_{-})/2$, $M_o=(M_{+}-M_{-})/2$)
\begin{eqnarray}
&A:& N_f>N_f^c, \theta=0;
M_e=0, M_o=0, M_+=0, M_-=0,
\nonumber\\
&B:& N_f<N_f^c, \theta=0;
 M_e>0, M_o=0, M_+=M_->0,
\nonumber\\
&C:& N_f<N_f^c, \theta\not=0;
 M_e>0, M_o>0, M_+>M_->0,
\nonumber\\
&D:& N_f<N_f^c, \theta=\theta_c(N_f);
(first \ order \ transition)
\nonumber\\
&E:& N_f>N_f^c, \theta\not=0;
 M_e=0, M_o=2M_+>0, M_+=-M_->0,
\end{eqnarray}
where the points $A, E$ are in the chiral symmetric phase
and $B, C, D$ are in the chiral-symmetry-breaking phase.
In the above we have considered the route starting at
$\theta=0$
($A \rightarrow B \rightarrow C \rightarrow D$) to justify
the linearization procedure adopted in this paper, since
$B_{\pm}(p)\equiv 0$ is a trivial solution at $\theta=0$.
It should be remarked that we can adopt another
linearization \cite{Miransky85} by replacing the
denominator $q^2+B^2(q)$ with $q^2+M^2$ where $M$
determines the normalization of the solution of the
linear SD equation, e.g., $M=B(0)$.
In \cite{KM94} and the subsequent paper we give the
thorough analysis on the choice of linearization
procedure which is compatible with the route of
approaching the critical point, especially,
$A \rightarrow E \rightarrow D$ starting at $\theta\not=0$
from the beggining where $B_{\pm}(p) \equiv 0$ is not a
trivial solution.


\section{Parity breaking}
We consider a problem whether or not the parity is dynamically
broken in (2+1)-dimensional QED, in other words, the
non-perturbatively  induced CS term may or may not break the parity.
When $\theta=0$, we obtain
\begin{eqnarray}
 D_T(k) =  {k^2+\alpha \sqrt{k^2} \over
(k^2+\alpha \sqrt{k^2})^2+\Pi_O(k)^2},
\end{eqnarray}
\begin{eqnarray}
 D_O(k) =  {\Pi_O(k) \over (k^2+\alpha \sqrt{k^2})^2+\Pi_O(k)^2}.
\end{eqnarray}
Here it should be remarked that the parity odd vacuum
polarization
$\Pi_O(k)$ is at least linearly proportional to $B_o(p)$,
since
\begin{eqnarray}
 \Pi_O(k) = 64 {\alpha \over \sqrt{k^2}} \int {d^3p \over
(2\pi)^3} (k^2 + p \cdot k) {B_o(p) \over (p+k)^2 p^2} +
{\cal O}(B_o^2).
\end{eqnarray}  Therefore the quantity $\Pi_O^2(k)$ in the
denominator of $D_T$ and $D_O$ should be neglected in the
linearized SD equation. This is sufficient to study the
critical phenomena in the neighborhood of the critical
point.
\par
{}From eq.~(\ref{SDBo}),
the linearized SD equation for the odd part $B_o$ reads
\begin{eqnarray}
 B_o(p) &=& m_o + {8\alpha \over N_f} \int {d^3q \over (2\pi)^3}
{B_o(q) \over q^2} \left[ 2 + \xi((p-q)^2) \right]
D_T(q-p)
\nonumber\\
&&- {16\alpha \over N_f} \int {d^3q \over (2\pi)^3}
{q \cdot (q-p) \over q^2\sqrt{(q-p)^2}} D_O(q-p),
\end{eqnarray}
where the non-local gauge is given by
$\xi(k^2)
= {2 \over 3} - {\sqrt{k^2} \over 3\alpha} + {\cal O}(k^2)$.
By mimicking the procedure of CCW \cite{CCW91}, the linearized SD
equation can be rewritten as
\begin{eqnarray}
 B_o(p) = m_o + {32 \over 3\pi^2 N_f} \int^\Lambda dq K(p,q) B_o(q),
\end{eqnarray}
with the kernel
\begin{eqnarray}
 K(p,q) &=& {p+q-|p-q| \over 2pq} - {9 \over 8\alpha}
- 6\pi \int^\Lambda dk {1 \over k^2}
\left({\alpha \over k+\alpha}\right)^2
\nonumber\\ && \times
\left[ 1+{k^2-p^2 \over 2pk} \ln {p+k \over |p-k|} \right]
\left[1+{k^2-q^2 \over 2qk} \ln {q+k \over |q-k|} \right].
\end{eqnarray}
This integral equation is too difficult to be solved analytically.
However it is quite easy to see that
the SD equation can not have the nodeless (ground state) solution
with a definite sign, since  the kernel $K(p,q)$ is always negative
for any  $p,q \in [0,\Lambda]$ as shown in Fig.~6.
This implies that the SD equation has only a trivial solution
$B_o(p) \equiv 0$.  Therefore the dynamical breakdown of parity
does not occur, which is in agreement with the previous analyses
\cite{ABKW86b,Poly88,RY86a,HM89}.


\section{Conclusion and Discussion}
 In this paper we have reexamined the dynamical breakdown
of chirality and parity in (2+1)-dimensional QED.  In order
to get rid of the ambiguity coming from the vertex
correction, we have derived the non-local gauge fixing
function which guarantees the absence of the wavefunction
renormalization for the fermion. Using the non-local gauge,
we have written down explicitly the SD equation for the
fermion mass function.
\par
 In the analyses done in this paper, we have
taken into  account the leading and next-to-leading terms
of the non-local gauge function  $\xi(k^2)$ in the IR
region to solve the SD equation analytically.  Within this
approximation we have found that there exists the finite
critical number of flavors $N_f^c=128/3\pi^2$ and that the
scaling form of the essential singularity type is obtained
in the absence of the CS term.
The critical value $N_f^c$ obtained in this paper reproduces
Nash's result \cite{Nash89} which is slightly larger than
the previous one
$32/\pi^2$ in the absence of the CS term $\theta =0$.
\par
In the presence of the CS term, it is shown \cite{KM94}
that the phase transition in the  direction of $\theta$
shows the discontinuity at some critical value,
$\theta_c(N_f)$, which depends on
$N_f$. The bare CS term decreases the critical flavour
number.
The order of chiral transition as well as the critical
line obtained in \cite{KM94} is completely different from
that of Hong and Park \cite{HP93}.
In Hong and Park \cite{HP93} the first order term of
$\theta$ was neglected in the SD equation for the fermion
mass function which corresponds to eq.~(\ref{SDBe}) and
hence this approximation inevitably forces the parity-odd
mass part to decouple completely from the SD equation for
the parity-even mass. This careless treatment leads to the
totally different conclusion: 1) the critical number of
flavors as a function of
$\theta$:
$ N_f^c(\theta) = N_f^c {1 \over 1+(\theta/\alpha)^2},
$ which implies the critical line
$ {\theta \over \alpha} = \sqrt{N_f^c/N_f-1}
$ where $N_f^c=N_f^c(\theta=0)$. 2) the continuous
(infinite order) phase transition characterized by the
essential singularity on the whole critical line:
$ {M_e \over \Lambda} \sim \exp \left[ {- 2n\pi \over
\sqrt{N_f^c(\theta)/N_f-1}} \right] ,
$ as in the absence of CS term. However this treatment is
not correct as was noticed in the footnote of the linear
approximation in section 2, since the CS term dominates
at large momentum and gives the essential contribution to
the UVBC which determines the character of the phase
transition. This fact was confirmed by the direct
numerical calculation of the non-linear SD equation
\cite{KM94} without relying on the specific linearization
procedure.

 \par
 In the same scheme we have shown that the dynamical
breakdown of  parity does not occur, which agrees with the
previous analyses
\cite{ABKW86b,Poly88,RY86a,HM89} and the general argument
\cite{VW84}.
  An interpretation of absence of dynamical parity
breaking is as follows.
 The gauge interaction acts as an attractive force to cause
the (chiral) condensation between fermion and antifermion,
$\langle \bar \psi \psi \rangle \not= 0$.
 As can be seen from the analysis in the previous section,
however, the induced CS term generates the same effect as
the repulsive force and this gets superior to the original
attractive force to result in the net repulsive force
(minus sign of the kernel) which prevents the
condensation,  $\langle \bar \psi \tau \psi \rangle = 0$.
Therefore it is necessary to incorporate other attractive
interactions so as to induce the parity  breakdown
$\langle \bar \psi \tau \psi \rangle \not= 0$, e.g., the
four-fermion interaction as done in CCW \cite{CCW91b}.

\par
 We must comment on the limitation of our approach.
 The explicit form of the non-local gauge we have found in
this paper can not be applied to the vertex which is not of
the form  $\Gamma_\mu(p,q) = \gamma_\mu F(p,q)$ where
$F(p,q)$ is the functional of $A$ and has the limit:
 $F(p,q) \rightarrow 1$ as $A(p) \rightarrow 1$.
 For the given vertex which has the different tensor
structure
\cite{CPW92},
therefore, we must rederive the corresponding non-local
gauge, if this is possible.
In such a case, we can not say anything about the result at
this stage.
 \par
In this paper we have truncated the $1/N_f$ series up to some finite
order and have taken into account the first two terms.
To improve our analyses, we should mention the effect of the
third term in the non-local gauge:
\begin{eqnarray}
 \xi(k^2) = {2 \over 3}  - {1 \over 3} {\alpha^2-\theta^2 \over
\alpha^2+\theta^2}  {\sqrt{k^2} \over \alpha}
+  {(\alpha^2-5\theta^2)(3\alpha^2+\theta^2) \over
15(\alpha^2+\theta^2)^2}
 {k^2 \over \alpha^2}
+ {\cal O}\left( \left({k \over \alpha}\right)^3 \right).
\end{eqnarray}
 The third term changes the equivalent differential equation
and hence the solution of the SD equation may change.
It may be expected that the qualitative feature of the
dynamical symmetry breaking will be unchanged.
 If the $1/N_f$ series can be summed as pointed out in \cite{CPW92},
however, the wavefunction renormalization effect can alter the
behavior of the dynamically important low momentum region,
which is recognized also in \cite{KN92}.
Moreover we have neglected all the terms of the order
${\cal O}(B_{\pm}^2)$ to obtain the closed SD equation for $B_e$
and $B_o$.  The resulting equation is a linear integral equation
where the IR cutoff must be introduced to define the equation
in the IR region.  This procedure may change the low momentum
behavior of the solution, in contrast to the UV region.
Anyway it is quite necessary to perform the numerical calculation
based on the non-local gauge without the expansion in order to
confirm our results.

 \par
Finally we mention another problem.
Though the inclusion of the bare CS term breaks the parity
explicitly, it is an interesting question whether the induced CS
term recovers the parity invariance which is explicitly broken at
the tree level, $\theta\not=0$.  In order to answer this question,
we must include both the bare and the induced CS terms, in which
case  it is difficult to solve the corresponding SD equation
analytically.   This problem will be discussed in a forthcoming
paper.

\par
One of the authors (K.-I. K.) is supported in part by the
Grant-in-Aid for  Scientific Research from the Ministry of
Education, Science and Culture (No. 05640339), and the International
Collaboration Program of the Japan Society for the Promotion of
Science.
We would like to thank Pieter Maris for
having pointed out the incomplete point in the first
version of this paper.  K.-I. K is grateful to Yoonbai
Kim for valuable comments on the effect of the Chern-Simons
term.

\newpage
\appendix
\section{Derivation of the non-local gauge}
In QED, we introduce the non-local gauge fixing
function $\xi(k^2)$ in the photon propagator:
\begin{equation}
  D_{\mu\nu}(k)
  = \frac{d(k^2)}{k^2}
    \left( g_{\mu\nu} - \eta(k^2) \frac{k_\mu
k_\nu}{k^2} \right),
\end{equation}
where
\begin{equation}
d(k^2) := k^2 D_T(k),
\ \eta(k^2) := 1-\xi(k^2).
\end{equation}
 In the euclidean QED$_D$ in D-dimensional space,
the SD equation for the fermion wave function
renormalization $A$ is given by
\begin{eqnarray}
 && p^2 A(p^2) -  p^2
\nonumber\\ &=& e^2 \int {d^Dq \over (2\pi)^D}
{A(q^2) \over q^2 A^2(q^2)+B^2(q^2)}
\nonumber\\ &&  \times d(k^2)
\left[ (D-2) {p \cdot q \over k^2} + \left( {p
\cdot q \over k^2} - 2 {p^2q^2-(p \cdot q)^2 \over
k^4} \right) \eta(k^2) \right],
\end{eqnarray}
and the SD equation for the fermion
mass function $B$ reads
\begin{eqnarray}
 B(p^2) &=& e^2 \int {d^Dq \over (2\pi)^D} {B(q^2)
\over q^2 A^2(q^2)+B^2(q^2)}
{d(k^2) \over k^2} [D-\eta(k^2)] .
\end{eqnarray}

Separating the angle $\vartheta$ defined by
\begin{equation}
  k^2 := (q-p)^2 = x + y -2\sqrt{xy} \cos\vartheta,
\  x := p^2, \qquad y := q^2,
\end{equation}
we find
\begin{eqnarray}
 && p^2 A(p^2) -  p^2
\nonumber\\ &=&   C_D e^2 \int_0^{\Lambda^2} dy {y
^{(D-2)/2}A(y) \over y A^2(y)+B^2(y)}
  \int_0^{\pi} d \vartheta \sin ^{D-2}
\vartheta
\nonumber\\ && \times d(k^2) \left[  {\sqrt{xy}
\cos \vartheta (D-2+\eta(k^2)) \over k^2}
 - 2 {xy-(\sqrt{xy} \cos \vartheta)^2 \over k^4}
\eta(k^2) \right],
\end{eqnarray} and
\begin{eqnarray}
 B(p^2) &=& C_D e^2  \int_0^{\Lambda^2} dy {y
^{(D-2)/2}B(y) \over y A^2(y)+B^2(y)}
 {d(k^2) \over k^2} [D-\eta(k^2)],
\end{eqnarray}
where
\begin{equation}
 C_D := {1 \over 2^D \pi^{(D+1)/2} \Gamma({D-1
\over 2})} .
\end{equation}

We perform the angular integration by parts as
\begin{equation}
  \int_0^\pi d\vartheta \sin^{D-2} \vartheta \cos\vartheta
\sqrt{xy} f(z)
  = \frac{\sqrt{xy}}{D-1} \left[ \sin^{D-1} \vartheta
f(z)
\right]_0^\pi
   -\frac{2xy}{D-1} \int_0^\pi d\vartheta \sin^D
\vartheta f'(z),
\end{equation}
where
\begin{equation}
 z := k^2 = (q-p)^2.
\end{equation}
Then we find
\begin{eqnarray}
  p^2 A(p^2) -  p^2 &=&
- e^2   C_D \int_0^{\Lambda^2} dy
{y ^{(D-2)/2}A(y) (2xy) \over y A^2(y)+B^2(y)}
 \int_0^{\pi} d \vartheta \sin ^{D} \vartheta
\nonumber\\ &&   \times
\left[ {1 \over D-1} {\partial \over \partial z}
\left( {(D-2+\eta(z))d(z) \over z} \right)
 +  {d(z) \eta(z) \over z^2}
\right]   .
\end{eqnarray}
 This is further rewritten as
\begin{eqnarray}
 p^2 A(p^2) -  p^2 &=&  - {C_D e^2 \over D-1}
\int_0^{\Lambda^2} dy {y ^{(D-2)/2}A(y) (2xy)
\over y A^2(y)+B^2(y)}
 \int_0^{\pi} d \vartheta \sin ^{D} \vartheta
\nonumber\\ && \times {1 \over z^{D-1}}
\left\{ (z^{D-2} d(z) \eta(z))' -(D-2)
z^{D-3}[d(z)-zd'(z)] \right\}  ,
\end{eqnarray}
where the prime denotes the differentiation with
respect to $z$.
\par
 The requirement $A(p^2) \equiv 1$
is achieved by taking $\eta(z)$ such that
$ (z^{D-2} d(z) \eta(z))' -(D-2)
z^{D-3}[d(z)-zd'(z)] \equiv 0$.
This is simply solved as follows.
\begin{eqnarray}
 \eta(z) &=& {D-2 \over z^{D-2} d(z)} \int_0^z dt
[d(t)-t d'(t)]t^{D-3} ,
\end{eqnarray}
where we have assumed that
$[z^{D-2} d(z) \eta(z)]|_{z=0}=0$ so as to
eliminate the
$1/z^{D-2}$ singularity in $\eta(z)$ and
$ \xi(z) = 1-\eta(z)$. This should be checked
after having obtained the function $\eta(z)$.
\par
 The gauge-fixing function $ \xi(z) =1-\eta(z)$
 is simplified as
\begin{eqnarray}
 \xi(z)  &=& D-1-{(D-1)(D-2) \over z^{D-2}d(z)}
\int_0^z dt  d(t) t^{D-3} ,
\label{nonlocalgauge}
\end{eqnarray}
where we have assumed
$[z^{D-2}d(z)]|_{z=0}=0$.
\par
For $D=3$, eq.~(\ref{nonlocalgauge}) reduces to
eq.~(\ref{nonlocalf}), since
$k^4D_T(k^2), k^4 D_T(k^2) \xi(k^2) \rightarrow 0$
as $k^2 \rightarrow 0$.

\section{Integration formulae}
Define the angle $\vartheta$  from
$k^2 :=(q-p)^2=p^2+q^2-2p q \cos \vartheta$,  where
$k=\sqrt{k^2}, p=\sqrt{p^2}, q=\sqrt{q^2}$. Then we obtain
the following results.
\begin{eqnarray}
\int_0^\pi d \vartheta \sin \vartheta
 {1 \over k}
 &=& {p+q-|p-q| \over pq} = {2 \over max(p,q)},
\nonumber\\
\int_0^\pi d \vartheta \sin \vartheta
 {\cos \vartheta \over k}
 &=& {(p+q)(p^2-pq+q^2)-|p-q|(p^2+pq+q^2) \over 3p^2q^2}
\nonumber\\
&=& {2 \over 3} {min(p,q) \over max(p^2,q^2)},
\nonumber\\
\int_0^\pi d \vartheta \sin \vartheta
 {1 \over k^2}
 &=& {1 \over pq} \ln {p+q \over |p-q|},
\nonumber\\
\int_0^\pi d \vartheta \sin \vartheta
 {\cos \vartheta \over k^2}
 &=& -{1 \over pq}
 +{p^2+q^2 \over 2p^2q^2} \ln {p+q \over |p-q|} .
\end{eqnarray}

\newpage

\centerline{{\large Figure Captions}}

\begin{enumerate}
\item[Fig.1:]
Schwinger-Dyson equation for the fermion propagator.

\item[Fig.2:]
Non-local gauge function $\xi(s)$ for $s:=k^2$
 (when $\theta=0$).
The lines from above to below correspond to
$\alpha/\Lambda=1, 10^{-1}, 10^{-2}, 10^{-4}$.

\item[Fig.3:]
Plot of the function $f(p)$ in
eq.~(\ref{functionf}), apart from the factor of the
integral .

\item[Fig.4:]
Schematic plot of the function $H_{\pm}(t)$
eq.~(\ref{alge2}) and
$\sin t$. The solution $M_{\pm}$ is obtained from the
intersection point of the curve $H_{\pm}(t)$ with $\sin t$.

\item[Fig.5:]
 Phase diagram (schematic).

\item[Fig.6:]
Integral kernel $K(p,q)$ as a function of $p$ and $q$ where
$\Lambda=\alpha$.

\end{enumerate}

\end{document}